E840 -- *De motu cometarum in orbitis parabolicis, solem in foco habentibus*
L. Evlero

On the motion of comets in parabolic orbits, having the Sun in the focus
L. Euler

Originally published in Opera Postuma 2, 1862, pp. 402-415[1]
Opera Omnia: Series 2, Volume 29, pp. 401 – 420

Translated and Annotated[2]
by
Sylvio R. Bistafa[*]
July 2020

Foreword

In this publication, Euler addresses six problems related to comets in heliocentric parabolic orbits. Problem 1: Find the true anomaly of a heliocentric comet from the latus rectum of the orbit and the medium Earth to Sun distance. Problem 2: Find the orbit of a heliocentric comet from three given positions. Problem 3: Knowing the orbit of a comet, and the instant in time in which it dwells in the perihelion, define its longitude and latitude at any time. Problem 4: From two locations of a heliocentric comet, find the inclination of the comet's orbit in relation to the ecliptic, and the positions of the nodes. Problem 5: Find the comet's true anomaly at any time before or after it had appeared in the perihelion. Problem 6: Find the orbit of a comet from three given heliocentric longitudes and latitudes. From these problems, several corollaries and scholia are derived by Euler.

_______________________

1. **Problem 1.** To find the motion of a comet in a given heliocentric parabolic orbit at a given time.

**Solution.** (Fig. 214) Let $MEAF$ be the parabolic orbit of a comet, in which focus $S$ dwells the Sun, and as so it is given, and it will be given the location of the perihelion $A$ in the sky as seen from the Sun, and also, the comet is seen from the Sun as moving in a great circle, which plane is equally given. Also, let be given the time at which the comet dwells in the perihelion, and in an earlier or later time, it will be sought the location of the comet $M$ as seen from the Sun.

---

[1] It was not possible to track down when this treatise was written, presumably around the 1740s.
[2] The Translator used the best of his abilities and knowledge to make this translation technically and grammatically as sound as possible. Nonetheless, interested readers are encouraged to submit suggestions for corrections as they see fitting.
[*] Corresponding address: sbistafa@usp.br

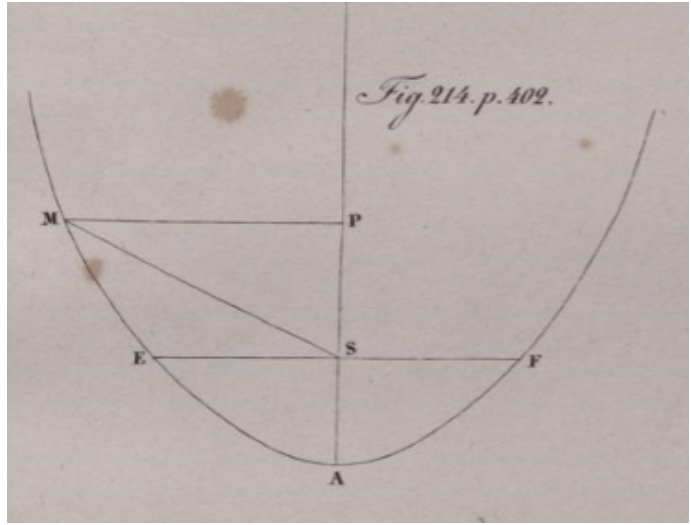

Fig. 214. p. 402.

Having elapsed the time $T$ after which the comet that appeared in the perihelion $A$ is, at this time, at the location $M$, such that the comet in the sky appears to stand apart the angle $ASM$ from the location $A$ of the perihelion, being this angle the true anomaly of the comet. Set this angle what is being looked for, $ASM = v$, and let be established that the distance from the perihelion to the Sun $SA = a$, and then the *latus rectum* parameter will be $= 4a$. From the location $M$ of the comet, it is drawn the perpendicular $MP$ to the axis of the parabola, and set $AP = x$, $PM = y$, and then $y^2 = 4ax$, and because $PS = x - a$, then the radius $SM = x + a$. Hence, the sine of the angle $ASM = v$ will be $= \frac{y}{a+x}$ and the cosine $= \frac{a-x}{a+x}$, assuming the whole sine $= 1$. Then, since $\cos v = \frac{a-x}{a+x}$, thus $x = \frac{a(1-\cos v)}{1+\cos v}$, and the distance from the comet to the Sun $MS = a + x = \frac{2a}{1+\cos v}$.

Therefore, having been found the true anomaly $v$, it becomes known the distance from the comet to the Sun $MS = \frac{2a}{1+\cos v}$. Since, truly, the time $T$, which the comet from the perihelion reaches the location $M$ is direct related to the area $ASM$, and inversely related to the square root of the latus rectum parameter $4a$, it is necessary to look for the area $ASM$ which is $=$ area $APM - \Delta SPM$. But, from the nature of the parabola, the area $APM$ is

$$= \frac{2}{3}xy, \text{ and } \Delta SPM = \frac{1}{2}y(x-a) = \frac{1}{2}xy - \frac{1}{2}ay;$$

whence the area $ASM$ will be $= \frac{1}{6}xy + \frac{1}{2}ay$. It is true that $x = \frac{a(1-\cos v)}{1+\cos v}$ and that $y = (a+x)\sin v = \frac{2a \sin v}{1+\cos v}$. Then, we have that $= \frac{1}{6}x + \frac{1}{2}a = \frac{2a+a\cos v}{3(1+\cos v)}$, therefore, the area $ASM = \frac{2a+a\cos v}{3(1+\cos v)} = \frac{2a^2(2+\cos v)\sin v}{3(1+\cos v)^2}$.

To simplify this expression, let us consider the tangent of half the angle $ASM$, or $\tan\frac{1}{2}v = t$, then $\sin\frac{1}{2}v = \frac{t}{\sqrt{1+t^2}}$, $\cos\frac{1}{2}v = \frac{1}{\sqrt{1+t^2}}$, thence $\sin v = \frac{2t}{1+t^2}$, $\cos v = \frac{1-t^2}{1+t^2}$, and further, $2 + \cos v = \frac{3+t^2}{1+t^2}$, and $1 + \cos v = \frac{2}{1+t^2}$. Finally,

$$\text{area } ASM = \frac{1}{3}a^2t(3+t^2) = a^2\left(t+\frac{1}{3}t^3\right).$$

Let now be considered that the major semi-axis of Earth's orbit, or the medium distance from the Earth to the Sun $= c$, and also, the Planet that revolves around the Sun with radius $= c$, had completed an absolute period of one sidereal year, that is $365^d 6^h 8'31''$, and let us put this time $= \theta$. Since the area of this circle is $\pi c^2$, denoting $1:\pi$ the ratio of the diameter to the circumference, and being the diameter parameter equal to $2c$, then the time $\theta$ of one revolution will be as the area $\pi c^2$ divided by $\sqrt{2c}$, that is, $\frac{\pi}{\sqrt{2}}c\sqrt{c}$. In a similar way, the time $T$ in which the comet coming from $A$ reaches $M$, will be as the area $ASM = a^2\left(t+\frac{1}{3}t^3\right)$ divided by $\sqrt{4a}$, that is $\left(t+\frac{1}{3}t^3\right)\frac{a\sqrt{a}}{2}$; whence arises this analogy $\theta:T = \frac{\pi}{\sqrt{2}}c\sqrt{c}:\frac{a\sqrt{a}}{2}\left(t+\frac{1}{3}t^3\right)$, then

$$t + \frac{1}{3}t^3 = \frac{\pi T c\sqrt{2c}}{\theta a\sqrt{a}} = 4.4428829381 \cdot \frac{T}{\theta} \cdot \frac{c\sqrt{c}}{a\sqrt{a}}.$$

Since $\theta$ is the sidereal year, and $T$ a given time, then $\theta$ to $T$ will be as $360°$ to the medium motion of the Earth convening to the time $T$. Then, if one considers that for the medium motion of the Earth $T$ is $= m$, then, $\frac{T}{\theta} = \frac{m}{360°}$. Then, from this cubic equation

$$t^3 + 3t = \frac{\pi T c\sqrt{2c}}{\theta a\sqrt{a}} = 13.3286488144 \cdot \frac{m}{360°} \cdot \frac{c\sqrt{c}}{a\sqrt{a}}$$

the value of $t$ itself is rooted out, which will be the tangent of half of the angle $ASM$, and then, at a given time before or after the passage of the comet by the perihelion $A$, the location of the comet can be assigned. **Q.E.I.**[3]

2. **Corollary 1.** If the time $T$ is the interval of one day or 24 hours, the $m = 59'8''$, whence, after calculation results in $t^3 + 3t = \frac{\pi T c\sqrt{2c}}{\theta a\sqrt{a}} = 0.036491289910 \cdot \frac{c\sqrt{c}}{a\sqrt{a}}$. Hence, if $T$ is the interval of $n$ days, then

$$t^3 + 3t = \frac{\pi T c\sqrt{2c}}{\theta a\sqrt{a}} = 0.036491289910 \cdot \frac{nc\sqrt{c}}{a\sqrt{a}}$$

or

$$t + \frac{1}{3}t^3 = \frac{\pi T c\sqrt{2c}}{\theta a\sqrt{a}} = 0.012163763303 \cdot \frac{nc\sqrt{c}}{a\sqrt{a}}.$$

3. **Corollary 2.** Then, for any number of days $n$, it becomes known the number itself equal to $t + \frac{1}{3}t^3$, whence, with difficulty, the proper value of $t$, from which, the value of the angle $v$ is found, then, it will be possible to construct a table, which, for each value of the angle $v$, exhibits the corresponding values of $t + \frac{1}{3}t^3$: also, in turn, with the assistance of this table, from a given value of $t + \frac{1}{3}t^3$, the angle $v$ will be obtained.

---

[3] *Quod Erat Inveniendum* => Which Was to Be Found Out

4. **Corollary 3.** Therefore, if a time is given, at which the comet dwells in the perihelion, and also the distance $a$ from the perihelion to the Sun, it is possible to determine the distance of the comet to the perihelion seen from the Sun with the aid of this table. From the number of days $n$ before or after the comet has arrived at the perihelion, the purpose is, of course, to compute the value of $0.012163763303 \cdot \frac{nc\sqrt{c}}{a\sqrt{a}}$; this value is entered in the table in the sub column $t + \frac{1}{3}t^3$, and the corresponding value of $v$ itself will give the sought angle.

5. **Example.** For the comet that appeared in the year 1680, Newton established the latus rectum of the orbit $4a = 236.8$, or $a = 59.2$ knowing that $c = 10000$, and also that this comet moved about the perihelion in December 8 1680 at $0^h 4'$p.m. Then, $0.012163763303 \cdot \frac{c\sqrt{c}}{a\sqrt{a}} = 26.70458$, and also $n$ days before or after the comet has arrived at the perihelion will give $t + \frac{1}{3}t^3 = 26.70458\, n$, from which the corresponding angle $v$ will give the distance of the comet to the perihelion in its orbit as seen from the Sun. Therefore, one day before or after reaching the perihelion this comet brought about an angle $ASM$ greater than $152°$; also, in the preceding or the anteceding day, it discloses approximately 6 degrees. Moreover, after the comet had passed the perihelion for an interval of as much as 90 days after having come into sight, then in this whole period it discloses an angle $ASM$ of around $174°$.

6. **Corollary 4.** Once the angle $ASM = v$ is found, it will be possible to know the distance $SM$ of the comet to the Sun, which is $= \frac{2a}{1+\cos v}$. Considering that $t = \tan\frac{1}{2}v$, then $\cos v = \frac{1-t^2}{1+t^2}$ and $1 + \cos v = \frac{2}{1+t^2}$. Then the distance $SM = a(1 + t^2) = a\, \sec^2\frac{1}{2}v$. Since the distance from the comet to the Sun revealed an angle $ASM = 174°$, where it began to disappear, on account of $a = 59.2$ and $t = \tan 87°$, then $SM = 21613$; showing that it didn't exceed very much the diameter $2c$ of the great orbit.

7. **Corollary 5.** If the tangent of the curve is taken from $M$, and in it the perpendicular is drawn from $S$, then, the length of this segment will be $= a\sqrt{(1+t^2)}$,[4] and the velocity of the comet in point $M$ will vary as $\frac{1}{a\sqrt{(1+t^2)}}$, and since $a$ is a constant, then it varies as $\cos\frac{1}{2}v$.[5]

8. **Scholium.** If the resulting number from $t + \frac{1}{3}t^3$ is not exactly found in the table, then, by the usual practice of interpolation the angle $v$ could be investigated in minutes and seconds, unless, by chance, that number is too big, and also when the angles resulting from the number $t + \frac{1}{3}t^3$ excessively disagree from an arithmetic progression. Then, in this case, a special device will be necessary that will seek from the nature of the progression, the exact determination of the angle $v$. As an example, the angle $ASM$ will be sought, which the comet of the Year 1680 had made in the time of 10 days: then, we will have that $t + \frac{1}{3}t^3 = 267.0458$, whence giving an angle $v$ between $167°$ and $168°$. Then, first, as established by the usual practice of interpolation:

---

[4] For this proof, one has to consider that the tangent line at any point of the parabola makes an angle with its axis which is equal to the angle between the tangent and its focal distance.
[5] It is not clear why this geometrical construction has to do with the velocity of the comet.

```
167°:    234,1492         267,0458
168°:    296,6044         234,1492
  60':    62,4552          32,8966
          1,0409            1,0409
                           32,8968
                            1,3158
                              296
```

then, the angle $v = 167°34'$ and $t = \tan 83°47'$. Let now be considered that $t = \tan(83°47' + m'')$, then,

```
t = 9,1802838 +  4145 m              lt  = 0,9628561 + 196 m
 257,8974000 + 350000 m              lt³ = 2,8885683 + 588 m
 267,0776838 + 354145 m              l3  = 0,4771213
                                     l⅓t³ = 2,4114470 + 588 m
                                     num. = 257,8974 + 350 m
```

therefore, $267.0776 + 0.0354\,m = 267.0458$, and $354\,m = -318$, hence, $m$ is even less than one minute, and then $v$ is truly equal to $167°34'$.

9. **Problem 1**. Find the orbit of a heliocentric comet from three given positions.

**Solution.** (Fig. 215) Let $LMNA$ be the orbit of the comet whose orbit is sought; and also the plane in which it [the orbit] becomes naturally apparent from two observations. Let be firstly observed the comet in the direction $SL$; secondly, in the direction SM, and thirdly, in the direction $SN$. Then the angles $LSM$ and $LSN$ are given, and, likewise, the differences in time among these observations. Set the time between the first and the second observation $= m$ days, between the first and the third $= n$ days. Further, set the tangent of half of the angle $LSM = f$, the tangent of half of the angle $LSN = g$. Set $z$ days as the time that it takes for the comet to go from $L$ to the perihelion $A$; then the time between the location $M$ and $A$ will be equal to $z - m$, and between the location $N$ and $A$ will be equal to $z - n$. Set the tangent of half of the angle $ASL = t$, then

$$\tan\tfrac{1}{2}ASM = \frac{t-f}{1+ft} \quad \text{and} \quad \tan\tfrac{1}{2}ASN = \frac{t-g}{1+gt}.$$

---

[6] Although it is not clear how this interpolation was accomplished, the angle found agrees with that obtained by solving the third degree polynomial.

[7] These operations seem to have been carried out to confirm the exactitude of the angle found by interpolation; however, it is not clear how they were accomplished.

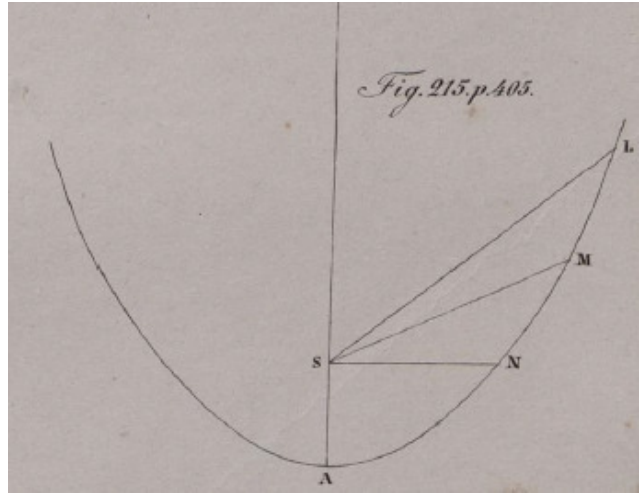

Fig. 215. p. 405.

Now, for the sake of brevity, set $N = 0.012163763303 \cdot \frac{nc\sqrt{c}}{a\sqrt{a}}$, the distance $SA = a$, and $c$ as the medium distance of the Earth to the Sun. Then, from these, we have that

$$t + \frac{1}{3}t^3 = Nz$$

$$\frac{t-f}{1+ft} + \frac{1}{3}\frac{(t-f)^3}{(1+ft)^3} = N(z-m)$$

$$\frac{t-g}{1+gt} + \frac{1}{3}\frac{(t-g)^3}{(1+gt)^3} = N(z-n)$$

and from these three equations, one should be able to determine the three unknowns $N$, $z$, and $t$. Subtracting the second and the third equation from the first equation, the following two equations are obtained

$$\frac{f(1+t^2)}{1+ft} + \frac{ft^2(1+t^2) - f^2 t(1-t^4) + \frac{1}{3}f^3(1+t^6)}{(1+ft)^3} = Nm$$

$$\frac{g(1+t^2)}{1+gt} + \frac{gt^2(1+t^2) - g^2 t(1-t^4) + \frac{1}{3}g^3(1+t^6)}{(1+gt)^3} = Nn$$

or

$$\frac{f(1+t^2)^2 + f^2 t(1+t^2)^2 + \frac{1}{3}f^3(1+t^2)^3}{(1+ft)^3} = Nm$$

$$\frac{g(1+t^2)^2 + g^2 t(1+t^2)^2 + \frac{1}{3}g^3(1+t^2)^3}{(1+gt)^3} = Nn$$

which by the division of one by the other gives an equation without the unknown $N$, and that involves only $t$

$$\frac{fn\left(1+\frac{1}{3}f^2+ft+\frac{1}{3}f^2t^2\right)}{(1+ft)^3} = \frac{gm\left(1+\frac{1}{3}g^2+gt+\frac{1}{3}g^2t^2\right)}{(1+gt)^3}$$

or

$$\frac{fn}{(1+ft)^2} + \frac{f^3n(1+t^2)}{3(1+ft)^3} = \frac{gm}{(1+gt)^2} + \frac{g^3m(1+t^2)}{3(1+gt)^3}$$

in this equation, the unknown $t$ now raises to the fifth power. For this reason, for the solution to become simpler, the three observations are chosen in such a way to be the least apart from each other, such that $f$ and $g$ turn out almost infinitely small. Then, truly, the first observation $L$ should not be far apart from the perihelion, such that $t$ can reduce the second term in both sides [of the equation] to negligible quantities[8]. Accordingly, the second term in both sides will vanish, and then

$$\frac{fn}{(1+ft)^2} = \frac{gm}{(1+gt)^2},$$

whence,

$$(1+gt)\sqrt{fn} = (1+ft)\sqrt{gm}, \quad \text{and then} \quad t = \frac{\sqrt{gm}-\sqrt{fn}}{g\sqrt{fn}-f\sqrt{gm}}.$$

Indeed in this way, the value of $t$ itself is found more near to the true [value], the less error had been committed in the observations, in large scale due to aberration. Then, if in this way, the value of $t$ itself is still found near to the true [value], then, any two other observations juxtaposed to the first [observation of] $L$, and although the equation, even being of the fifth power, eventually, and given now that the approximate true value of $t$ is known, its real value can be obtained without difficulty. If $\theta$ is the approximate value of $t$ itself, set $t = \theta + \psi$, such that $\psi$ is very small when compared to $\theta$, then

$$\frac{fn}{(1+f\theta)^2} + \frac{f^3n(1+\theta\theta)}{3(1+f\theta)^3} - \frac{2ffn\psi}{(1+f\theta)^3} - \frac{f^3n\psi(3f-2\theta+f\theta\theta)}{3(1+f\theta)^4} =$$

$$\frac{gm}{(1+g\theta)^2} + \frac{g^3m(1+\theta\theta)}{3(1+g\theta)^3} - \frac{2ggm\psi}{(1+g\theta)^3} - \frac{g^3m\psi(3g-2\theta+g\theta\theta)}{3(1+g\theta)^4}$$

[9]

once the value of $\psi$ is obtained, it then gives the true value of the tangent as $t = \theta + \psi$, and twice of the corresponding angle itself will indicate the angle $LSA$, and then, it will show the sought position of the parabola axis $AS$. Moreover, once $t$ is found, then

$$N = \frac{(1+t^2)^2\left[f+f^2t+\frac{1}{3}f^3(1+t^2)\right]}{m(1+ft)^3} = 0.0121637 \cdot \frac{nc\sqrt{c}}{a\sqrt{a}}$$

---

[8] Initially, $t$ was associated with the angle $ASM$ (Fig. 214), and now $t$ seems to be associated with the angle $ASL$ (Fig. 215).
[9] The origin of this expression is not clear.

whence, the distance $AS = a$ is brought forth, as well as the *latus rectum* parameter $4a$. Next, the time which the comet will take to reach the perihelion is acquired, which after the first observation in $L$ will happen in $z$ days, knowing that $z = \frac{3t+t^3}{3N}$. Hence, knowing the position of the parabola axis $AS$, its parameter $4a$, or the value of $N$, along with the time that it took for the comet to reach the perihelion, at any time of the comet orbit, its distance to the sun is defined by the preceding problem. **Q. E. I.** (*Quod Erat Inveniendum* => Which Was to Be Found Out).

10. **Corollary. 1.** From the knowledge of the time in which the comet approaches the perihelion, together with the value of the number $N$, in a given time, the distance of the comet to the perihelion as seen from the Sun can be determined; certainly if this distance is desired in $n$ days before or after it [the comet] having been passed the perihelion, the number $N$ should be multiplied by $n$, and the result will be entered in the table in the sub column $t + \frac{1}{3}t^3$, which will correspond to the angle $v$, revealing the distance of the comet to the perihelion as seen from the Sun.

11. **Corollary. 2.** When the value of the number $N$ that was found, is divided by 0.01216376330, the result will give $\frac{nc\sqrt{c}}{a\sqrt{a}}$, whence, the distance of the perihelion to the Sun $SA = a$ will be found, or else, its ratio to the medium distance of the Earth to the Sun.

12. **Scholion.** The main posed difficulty, is to find the tangent $t$ from the equation

$$\frac{fn}{(1+ft)^2} + \frac{f^3n(1+t^2)}{3(1+ft)^3} = \frac{gm}{(1+gt)^2} + \frac{g^3m(1+t^2)}{3(1+gt)^3}$$

where we first instruct to make the observations of $f$ and $g$ such that they are rather small, so that $t$ can only be approximated. But since the smallest error in the observations can produce a conclusion that considerable deviates from the truth, it will be convenient to take two additional observations, both not too near, and also not too far apart from the first, such that $f$ and $g$ are not too small, neither approaching unit, and this will be accomplished as long as the angle $LSN$ is less than $60°$. If such observations are adopted, then $\frac{fn}{(1+ft)^2} = \frac{gm}{(1+gt)^2}$ will give a value of $t$ itself, in fact deviating from the true value, but not too much. Set $t = \theta$, such that $\frac{fn}{(1+f\theta)^2} = \frac{gm}{(1+g\theta)^2}$, and let $t = \theta + \psi$, where $\psi$ will be considered a very small counterpart quantity. We will then consider using it as a solution of the following equation

$$\frac{f^3n(1+\theta\theta)}{3(1+f\theta)^3} - \frac{f^3n\psi(6+3f\theta+4f\theta+f\theta\theta)}{3(1+f\theta)^4} = \frac{g^3m(1+\theta\theta)}{3(1+g\theta)^3} - \frac{ggm\psi(6+3gg+4g\theta+gg\theta\theta)}{3(1+g\theta)^4}$$

which, on account of $fn : gm = (1+f\theta)^2 : (1+g\theta)^2$, it is transformed into

$$\frac{ff(1+\theta\theta)}{1+f\theta} - \frac{f\psi(6+3ff+4f\theta+ff\theta\theta)}{(1+f\theta)^2} = \frac{gg(1+\theta\theta)}{1+g\theta} - \frac{g\psi(6+3gg+4g\theta+gg\theta\theta)}{(1+g\theta)^2}$$ [10]

once $\psi$ is obtained from this equation, a sufficient close value of $t$ will be obtained as $t = \theta + \psi$, which, nevertheless, can be further corrected in a similar way. Finally, a resolution will require applying three observations furthest away from each other, and by an equation of fifth degree, from which the value of $t$ itself is determined with maximum exactitude, that which will be provided without difficult, when the approximate value of $t$ itself is known.

13. **Problem 3**. Knowing the orbit of a comet, together with the instant in time in which it dwells in the perihelion, define the longitude and the latitude of the heliocentric comet at any time.

**Solution.** (Fig. 216) The comet moves around the Sun in a plane that coincides with a great circle[11]. Let the Sun $S$ be located in the center of the constellation, such that ☊$MA$♈ is a great circle, in which the comet is seen to engage, according to the sequence of letters ☊$MA$♈. Set further ☊$ma$♈ as the ecliptic configured according to the sequence of symbols, and $O$ the northern ecliptic pole, then ☊ will be the ascending node of the comet's orbit, and ♈ its descending node. Set $q$ as the longitude of the ascending node ☊ measured from the first star of Aries, and the inclination of the comet's orbit ☊$MA$♈ in relation to the ecliptic, or angle $M$☊$m = s$, which if it is less than a right angle, the motion of the comet will be seen to take place according to the sequence of symbols; and conversely, if the angle $s$ is greater than a right angle, the motion of the comet will be executed contrary to the sequence of symbols. Next, set the location of the perihelion in $A$, through which the great circle $OAa$ is drawn from $O$, $a$ will be the longitude of the perihelion, whose distance to the first star of Aries is $= p$, and then, the arc ☊$a = p - q$. From the spherical triangle $A$☊$a$, with a right angle in $a$, the northern latitude of he perihelion $Aa$ becomes first known, and then, certainly, $\tan Aa = \tan s \cdot \sin(p - q)$ and $\tan ☊A = \frac{\tan(p-q)}{\cos s}$.[12]

---

[10] The origin of these expressions is not clear.
[11] A great circle (also known as an orthodrome) of a sphere is the intersection of the sphere and a plane that passes through the center point of the sphere. A great circle is the largest circle that can be drawn on any given sphere.
[12] These relations come from the application of Napier's rules for this right spherical triangle.

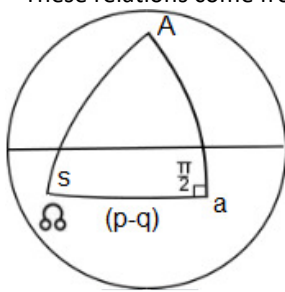

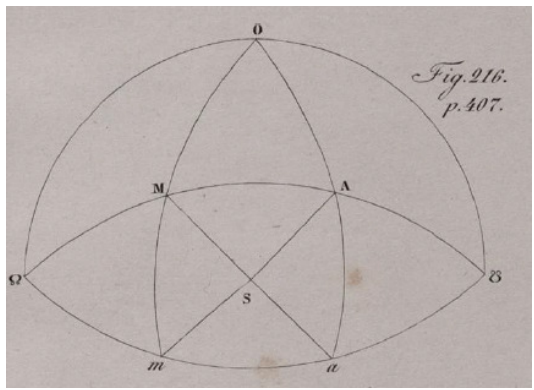

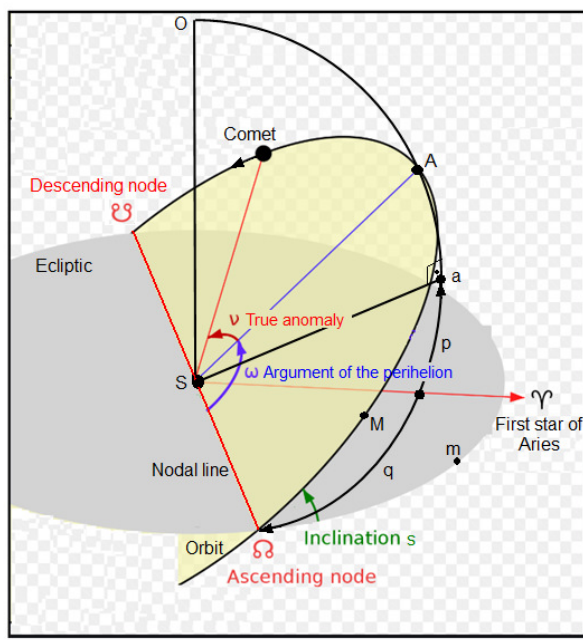

[13]

Set $a$ as the distance of the comet at the perihelion to the Sun, and $c$ the medium distance of the Earth to the Sun, and set

$$N = 0.012163763303 \cdot \frac{nc\sqrt{c}}{a\sqrt{a}}$$

Now, for a comet that is in $M$, it will be sought the number of days $k$ that it takes to reach the perihelion. To which is to be found the angle $MSA$ or the arc $MA = v$ and $tan\frac{1}{2}v = t$, and then $t + \frac{1}{3}t^3 = Nk$, and with the assistance of the calculation table, the angle $v$ will be found from the number $Nk$. Set the arc $\mathrm{\Omega} A = A$, such that $tan\, A = \frac{tan(p-q)}{cos\, s}$; then the arc $\mathrm{\Omega} M = A - v$. Hence, in the triangle $\mathrm{\Omega} mM$, with a right angle in $m$, we have that $sin\, Mm = sin(A - v) \cdot$ $\sin s$ and $\tan ☊m = \tan(A − v) \cdot \cos s$.[14] Therefore, $Mm$ will be the latitude of the heliocentric comet. And, if to ☊$m$ the longitude $q$ of the node is added, then the computed longitude of the comet in relation to the first star of Aires will come forth. **Q. E. I.**

---

[13] This figure was added by the Translator.

**Corollary. 1.** In the spherical triangle $M☊m$, the sides $Mm$ and the angle $M☊m = s$ are mutually dependent, such that $\tan Mm = \sin ☊m \cdot \tan s$. And this equation should result from the two just found, if the arc $A − v$ is eliminated between them.

14. **Corollary 2.** But if from the two equations found namely: $\sin Mm = \sin(A − v) \cdot \sin s$ and $\tan ☊m = \tan(A − v) \cdot \cos s$, the angle $s$ is eliminated between them, the following equation will appear

$$\cos(A − v) = \cos Mm \cdot \cos ☊m.$$

15. **Corollary 3.** The following four equations are then applicable, which are equivalent to the two initially found, these are

$$\sin Mm = \sin(A − v) \cdot \sin s, \qquad \tan ☊m = \tan(A − v) \cdot \cos s$$
$$\tan Mm = \sin ☊m \cdot \tan s, \qquad \cos(A − v) = \cos Mm \cdot \cos ☊m$$

two of these [relations], which are seen as the most convenient, will be used.

17. **Corollary 4.** Therefore, from these, the longitude and the latitude of the heliocentric comet at a given assign time, are first found from the orbit of the comet. Then, the distance of the node ☊ to the perihelion $A$ as seen from the Sun $= A$; and also, the inclination of the comet's orbit in relation to the ecliptic plane, or angle $M☊m = s$. Further, the difference between the proposed time and the time in which the comet touches the perihelion, which is expressed as $= k$ days, from which, with the aid of the table for the given data, gives the angle $v$.

18. **Corollary 5.** Since the distance of the comet to the Sun $SM$ is $= a \sec^2 \frac{1}{2} v$, then its shortened distance to the Sun, or that taken on the ecliptic $= a \sec^2 \frac{1}{2} v \cdot \cos Mm$.

19. **Corollary 6.** If the heliocentric longitude is set $= f$, and the heliocentric latitude $= g$,[15] then, $f = ☊m + q$, and then

$$I. \sin g = \sin(A − v) \cdot \sin s, \qquad II. \tan(f − q) = \tan(A − v) \cdot \cos s$$

---

[14] These relations come from the application of Napier's rules for this right spherical triangle.

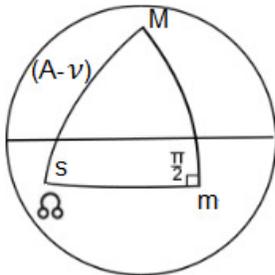

[15] Note that these same letters $f$ and $g$ were assigned to other geometrical elements of the comet's orbit in Problem1.

$$III. \tan g = \sin(f - q) \cdot \tan s, \qquad IV. \cos(A - v) = \cos g \cdot \cos ☊m.$$

20. **Problem 4.** From two locations of a heliocentric comet, find the inclination of the comet's orbit in relation to the ecliptic, and the positions of the nodes ☊ and ☋.

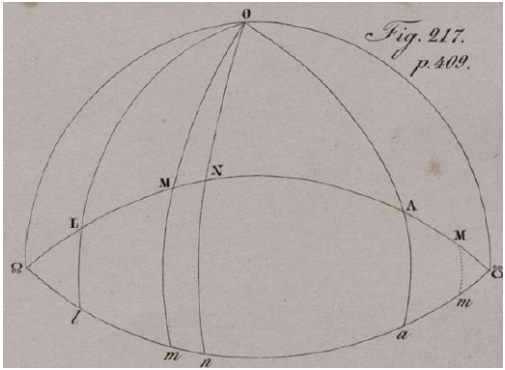

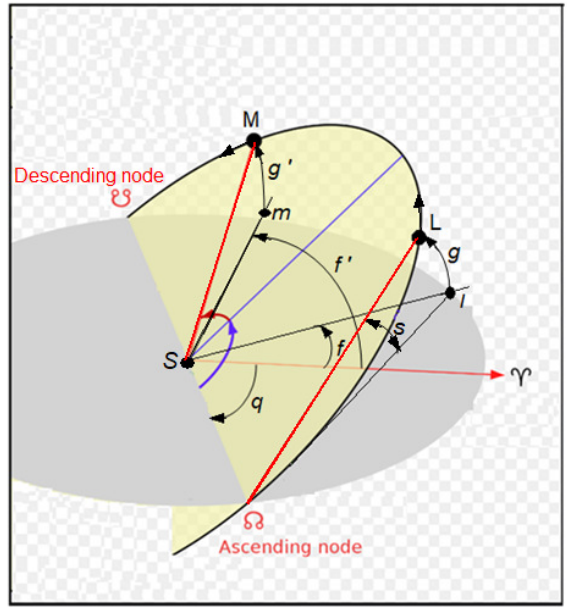

[16]

**Solution.** The comet is first observed in $L$, and let $f$ be the heliocentric longitude and $g$ the northern latitude; thereafter, the comet is then observed in $M$, and let $f'$ be its longitude and $g' = Mm$ its latitude. Set $q$ as the longitude of the ascending node ☊, and the orbit's inclination to the ecliptic, or angle $L☊l = s$. Then the difference of longitudes $lm = f' - f$. From the third equation (19), the following two equations will appear

$$\tan g = \sin(f - q) \cdot \tan s \quad \text{and} \quad \tan g' = \sin(f' - q) \cdot \tan s$$

and dividing one by the other gives

$$\frac{\tan g'}{\tan g} = \frac{\sin(f' - q)}{\sin(f - q)} = \frac{\sin(f - q + lm)}{\sin(f - q)}.$$

---

[16] This figure was added by the Translator.

But since $sin(f - q + lm) = sin(f - qm) \cos lm + \cos(f - q) \sin lm$, then

$$\frac{\tan g'}{\tan g} = \cos lm + \frac{\sin lm}{\tan(f - q)}.$$

to find out that

$$\tan(f - q) = \frac{\tan g \sin lm}{\tan g' - \tan g \cos lm}.$$

Then, the difference in longitude $☊l = f - q$ becomes known, whence, once the longitude of the point $l$ is given, the longitude of the ascending node will be known; once the arc f-q is known, then $\tan s = \frac{\tan g}{\sin(f-q)}$, and then becomes known the inclination of the comet's orbit to the ecliptic, since the angle $L☊l = s$. **Q. E. I.**

21. **Corollary 1.** Since $\tan(f - q) = \frac{\tan f - \tan q}{1 + \tan f \tan q}$ and $lm = f' - f$, then, the desired longitude of the node will be

$$\tan q = \frac{\sin f \tan g' - \sin f' \tan g}{\cos f \tan g' - \cos f' \tan g}.$$

22. **Corollary 2.**

$$\tan s = \frac{\sqrt{\alpha^2 + \beta^2 - 2 \alpha \beta \mu}}{\mu},$$

whose numerator of the fraction $\sqrt{\alpha^2 + \beta^2 - 2 \alpha \beta \mu}$ is the base of the triangle whose sides are $\alpha$ and $\beta$, and an angle $= f' - f$ between them.

23. **Problem 5.** If a given time, say $k$ days before or after the comet have reached the perihelion, and given that the distance[17] of the comet to the perihelion as seen from the Sun $= v$, find the same distance in $k + x$ days time before or after the time in which it appears in the perihelion.

**Solution.** Set $\tan \frac{1}{2}v = t$, then $t + \frac{1}{3}t^3 = Nk$, or $\tan \frac{1}{2}v + \frac{1}{3}\tan^3 \frac{1}{2}v = Nk$. Set now $k + x$ days to correspond to an angle $= v + \varphi$, and for the sake of brevity let us write $\tan \frac{1}{2}v + \frac{1}{3}\tan^3 \frac{1}{2}v = V$, and from the calculus of finite differences we may write

$$\tan \frac{1}{2}(v + \varphi) + \frac{1}{3}\tan^3 \frac{1}{2}(v + \varphi) = V + \varphi \frac{dV}{dv} + \varphi^2 \frac{d^2V}{2dv^2} + \varphi^3 \frac{d^3V}{6dv^3} + etc. = N(k + x).$$

Since $V = Nk$, then

$$Nx = \varphi \frac{dV}{dv} + \varphi^2 \frac{d^2V}{2d^2v} + \varphi^3 \frac{d^3V}{6dv^3} + etc.$$

Yet,

---
[17] This is not a distance, but actually the angle $ASM = v$ (see Fig. 214 in Problem 1.)

$$\frac{dV}{dv} = \frac{1}{2\cos^2\frac{1}{2}v} + \frac{\tan^2\frac{1}{2}v}{2\cos^2\frac{1}{2}v} = \frac{1}{2\cos^4\frac{1}{2}v}$$

$$\frac{d^2V}{dv^2} = \frac{\sin\frac{1}{2}v}{\cos^5\frac{1}{2}v}$$

$$\frac{d^3V}{dv^3} = \frac{1}{2\cos^4\frac{1}{2}v} + \frac{5\sin^2\frac{1}{2}v}{2\cos^6\frac{1}{2}v} = \frac{5}{2\cos^6\frac{1}{2}v} - \frac{2}{\cos^4\frac{1}{2}v}$$

etc.

And from these the result is

$$Nx = \frac{\varphi}{2\cos^4\frac{1}{2}v} + \frac{\varphi^2 \sin\frac{1}{2}v}{2\cos^5\frac{1}{2}v} + \frac{5\varphi^3}{12\cos^6\frac{1}{2}v} - \frac{\varphi^3}{3\cos^4\frac{1}{2}v} + etc.$$

Let us now set $\varphi = \alpha Nx + \beta N^2 x^2 + \gamma N^3 x^3 + etc.$, and once this substitution is made we will have that

$$Nx = \frac{\alpha Nx}{2\cos^4\frac{1}{2}v} + \frac{\beta N^2 x^2}{2\cos^4\frac{1}{2}v} + \frac{\gamma N^3 x^3}{2\cos^4\frac{1}{2}v} + etc.$$

$$+ \frac{\alpha^2 N^2 x^2 \sin\frac{1}{2}v}{2\cos^5\frac{1}{2}v} + \frac{\alpha \beta N^3 x^3 \sin\frac{1}{2}v}{\cos^5\frac{1}{2}v} + etc.$$

$$\frac{5\alpha^3 N^3 x^3}{12\cos^6\frac{1}{2}v}$$

$$-\frac{\alpha^3 N^3 x^3}{3\cos^4\frac{1}{2}v}.$$

These will be the result of the equality

$$\alpha = 2\cos^4\frac{1}{2}v$$

$$\beta = \frac{-\alpha^2 \sin\frac{1}{2}v}{\cos\frac{1}{2}v} = -4\cos^7\frac{1}{2}v \sin\frac{1}{2}v$$

$$\gamma = \frac{-2\alpha\beta \sin\frac{1}{2}v}{\cos\frac{1}{2}v} - \frac{5\alpha^3}{6\cos^2\frac{1}{2}v} - \frac{2\alpha^3}{3}, \text{ or } \gamma = \frac{4}{3}\cos^{10}\frac{1}{2}v\left(7 - 8\cos^2\frac{1}{2}v\right)$$

etc.

From these, it is found that

$$\varphi = 2Nx\cos^4\frac{1}{2}v - 4N^2x^2\cos^7\frac{1}{2}v\sin\frac{1}{2}v + \frac{4}{3}N^3x^3\cos^{10}\frac{1}{2}v\left(7 - 8\cos^2\frac{1}{2}v\right)$$

this expression gives a satisfactory approximation to the value of $\varphi$ itself, provided the difference in time $x$ is not very long. First, if the comet is very distant from the perihelion, the angle $\frac{1}{2}v$ will not defer too much from a right angle, such that its cosine fraction will be rather small. Then, the second term will be much less than the first, and the third to the second; then, generally, the first term is sufficient to give the value of $\varphi$, which once found, the angle sought will be $= v + \varphi$. **Q. E. I.**

24. **Corollary 1.** Then, if the comet is moved away from the perihelion by an angle $v$ during $k$ days, during the time of $k + x$ days, it will be moved approximately by the angle $v + 2Nx\cos^4\frac{1}{2}v$; and if this angle is desired almost exact, it will be $v + 2Nx\cos^4\frac{1}{2}v - 4N^2x^2\cos^7\frac{1}{2}v\sin\frac{1}{2}v$, and also the third term

$$\frac{4}{3}N^3x^3\cos^{10}\frac{1}{2}v\left(7 - 8\cos^2\frac{1}{2}v\right)$$

can always be safely ignored in the sequence, provided in the other time the comet had moved either before or after the perihelion zone.

25. **Scholium 1.** For this approximation to be further confirmed, let us consider, for example, the comet that appeared in the Year 1680, for which $N = 26.70458$, and since this number is much greater than unit, the terms of its series, exhibiting the value of $\varphi$ itself will be quite evident. Since this comet is moved around the Sun in one day by an angle greater than 152° from the perihelion, this comet can be considered as been observed long before or after it has been passed the perihelion, [as per 5.] $v > 152°$ and $\cos\frac{1}{2}v > \sin 14°$, and then $\cos\frac{1}{2}v > 0{,}2419219$, that is, $< \frac{1}{4}$. Hence, $\cos^4\frac{1}{2}v < \frac{1}{256}$, wherefore the value of the term $2Nx\cos^4\frac{1}{2}v$ is rendered quite small. Let us assume a time interval $k$ equal to ten days, and then [according to 8.] $v = 167°34'$ and $\frac{1}{2}v = 83°47'$. And if now, from these results, we look for the angle that corresponds to a time of $k + x$ days, the following calculation is set up

```
l cos ½ v = (— 1),0345825        lN  =   1,4265857
l sin ½ v = (— 1),9974386        l2  =   0,3010300
        lN =    1,4265857        l cos⁴ ½ v = (—4),1383300
                                 ─────────────────────
                                          (—3),8659457
```
[18]

---

[18] It is not clear how these operations are carried out. Nonetheless, it appears that these numerical results were obtained from tables of logarithms and trigonometric functions.

then, $2N \cos^4 \frac{1}{2} v = 0.007344$.

For the second term we will have that

$$lN^2 = 2{,}8531714$$
$$l4 = 0{,}6020600$$
$$l \cos^7 \tfrac{1}{2} v = (-7){,}2420775$$
$$l \sin \tfrac{1}{2} v = (-1){,}9974386$$
$$\overline{(-4){,}6947475}$$

then, $4N^2 \cos^7 \frac{1}{2} v \sin \frac{1}{2} v = 0.000495$.

Each expression of the third member is compute separately, namely

$$lN^3 = 4{,}2797571 \qquad\qquad lN^3 = 4{,}2797571$$
$$l \tfrac{4}{3} = 0{,}1249387 \qquad\qquad l \tfrac{4}{3} = 0{,}1249387$$
$$l7 = 0{,}8450980 \qquad\qquad l8 = 0{,}9030900$$
$$l \cos^{10} \tfrac{1}{2} v = (-10){,}3458250 \qquad l \cos^{12} \tfrac{1}{2} v = (-12){,}4149900$$
$$\overline{(-5){,}5956188} \qquad\qquad \overline{(-7){,}7227758}$$

then, $\frac{28}{3} N^3 \cos^{10} \frac{1}{2} v = 0.00003941$ and $\frac{42}{3} N^3 \cos^{12} \frac{1}{2} v = 0.0000005282$.

Then, in a time span of $10 + x$ days, the comet is moved away from the perihelion by an angle of

$$167°34' + 0.007344x - 0{,}0004839x^2 + 0.00003888x^3$$

whose terms, unless $x$ surpasses ten days, decreases remarkably. Set the time interval of one day, then

$$+ 0{,}007344$$
$$- 0{,}000484$$
$$+ 0{,}000039$$
$$\overline{\varphi = 0{,}006899} \qquad \sin 23' = 0{,}006690$$
$$0{,}006690 \qquad\qquad \sin 24' = 0{,}006981$$
$$\overline{\qquad\qquad\qquad 291}$$
$$291 : 60'' = 209 : 43''$$

[19]

---

[19] Since $\varphi = 0.006899\ rad \Rightarrow 0.3953° \Rightarrow 23'43''$, it is not clear why this more complicated procedure was adopted to find the angle $\varphi$.

hence, the time of eleven days corresponds to an angle of $167°57'43''$.

26. **Scholium 2.** If we put $x$ negative, then all the terms of the series exhibiting the value of $\varphi$ itself with the same signs will be affected, and, therefore, the series converges more. But, assuming that the time of $k$ days corresponds to an angle $v$ from the perihelion, or true anomaly, the time of $k - x$ days will correspond to a true anomaly of $v - \varphi$, such that

$$\varphi = 2Nx \cos^4 \frac{1}{2}v + 4N^2x^2 \cos^7 \frac{1}{2}v \sin \frac{1}{2}v + \frac{4}{3}N^3x^3 \cos^{10}\frac{1}{2}v \left(7 - 8\cos^2 \frac{1}{2}v\right)$$

and as we first saw, converges strongly, if the angle $\frac{1}{2}v$ falls short from a right angle, that is, if the comet is still far off the perihelion, even though in this case, $N$ is a rather big number. But if $N$ is a much smaller number, which occurs if the perihelion of the comet to the Sun is remoter, then this series converges satisfactorily, even if the comet is not too far from the perihelion.

27. **Problem 6.** Find the orbit of a comet from three given heliocentric longitudes and latitudes.

**Solution.** Set the longitude of the perihelion $= p$, the distance of the perihelion to the Sun $= a$, the medium distance of the Earth to the Sun $= c$, and be taken $N = 0.012163763303 \cdot \frac{c\sqrt{c}}{a\sqrt{a}}$. Set the longitude of the ascending node $= q$, the inclination of the comet's orbit to the ecliptic $= s$; and be introduced $r$, such that $\tan r = \frac{\tan(p-q)}{\cos s}$,[20] and then $r$ will be the distance of the perihelion to the node. Be considered three observations that have been taken long before the comet reaches the perihelion, or as soon as it begins to appear. Set for the observation

|  | I. | II. | III. |
|---|---|---|---|
| heliocentric longitude of the comet $=$ | $f$ | $f'$ | $f''$ |
| heliocentric longitude of the comet $=$ | $g$ | $g'$ | $g''$ |

time between observation $I$. and $II. = x$ days between $I$. and $III. = \lambda$ days.

After the first observation, set the time interval for the comet to reach the perihelion as $k$ days, and set the true anomaly of the comet at the time of the first observation $= v$, for the second $= v - \varphi$, and for the third $= v - \psi$, so that as we saw before,

$$\varphi = 2Nx \cos^4 \frac{1}{2}v + 4N^2x^2 \cos^7 \frac{1}{2}v \sin \frac{1}{2}v + \frac{4}{3}N^3x^3 \cos^{10}\frac{1}{2}v \left(7 - 8\cos^2 \frac{1}{2}v\right) + etc.$$

$$\psi = 2N\lambda \cos^4 \frac{1}{2}v + 4N^2\lambda^2 \cos^7 \frac{1}{2}v \sin \frac{1}{2}v + \frac{4}{3}N^3\lambda^3 \cos^{10}\frac{1}{2}v \left(7 - 8\cos^2 \frac{1}{2}v\right) + etc.$$

Now, as per (19), with $r$ being substituted for $A$

1) $\sin g = \sin(r - v) \cdot \sin s$       2) $\tan(f - q) = \tan(r - v) \cdot \cos s$

   $\sin g' = \sin(r - v + \varphi) \cdot \sin s$       $\tan(f' - q) = \tan(r - v + \varphi) \cdot \cos s$

   $\sin g'' = \sin(r - v + \psi) \cdot \sin s$       $\tan(f'' - q) = \tan(r - v + \psi) \cdot \cos s$

---

[20] This relation comes from $\tan ☊A = \frac{\tan(p-q)}{\cos s}$, where $r$ has been substituted for $☊A$ (see footnote 10).

3) $tan\, g = sin(f - q) \cdot tan\, s$       4) $cos(r - v) = cos\, g \cdot cos(f - q)$

   $tan\, g' = sin(f' - q) \cdot tan\, s$       $cos(r - v + \varphi) = cos\, g' \cdot cos(f' - q)$

   $tan\, g'' = sin(f'' - q) \cdot tan\, s$       $cos(r - v + \psi) = cos\, g'' \cdot cos(f'' - q)$

From equations 3) we have the following

$$\frac{sin(f' - q)}{sin(f - q)} = \frac{tan\, g'}{tan\, g} = \frac{sin\, f' cos\, q - cos\, f' sin\, q}{sin\, f\, cos\, q - cos\, f\, sin\, q}$$

and thence

$$tan\, q = \frac{tan\, g' sin\, f - tan\, g\, sin\, f'}{tan\, g' cos\, f - tan\, g\, cos\, f}.$$

The same value for the longitude of the node $q$ should result from other two equations of the same class, according to the observations that are instituted with maximum care; then the following will be equally applicable

$$tan\, q = \frac{tan\, g'' sin\, f' - tan\, g' sin\, f''}{tan\, g'' cos\, f' - tan\, g' cos\, f''}.$$

Then, once the longitude of the ascending node has been found, the inclination of the comet's orbit to the ecliptic $s$ becomes simultaneously known from the equation $tan\, s = \frac{tan\, g}{sin(f-q)}$. Further, because the angles $\varphi$ and $\psi$ are rather small, then $sin(r - v + \varphi) = sin(r - v) + \varphi\, cos(r - v)$ and $sin(r - v + \psi) = sin(r - v) + \psi\, cos(r - v)$, whence, from the equations of the first class it is found that

$$\frac{sin\, g'}{sin\, g} = 1 + \varphi\, cot(r - v) \quad \text{and} \quad \frac{sin\, g''}{sin\, g} = 1 + \psi\, cot(r - v)$$

whence

$$\frac{sin\, g'' - sin\, g}{sin\, g' - sin\, g} = \frac{\psi}{\varphi} = \frac{\lambda + 2N\lambda^2 cos^3 \frac{1}{2}v\, sin\frac{1}{2}v}{x + 2Nx^2 cos^3 \frac{1}{2}v\, sin\frac{1}{2}v}$$

and since $sin(r - v) = \frac{sin\, g}{sin\, s}$, the $cot(r - v)$ will also be known, whence, we will have that

$$\frac{sin\, g' - sin\, g}{sin\, g\, cot(r - v)} = \varphi = 2Nx\, cos^4 \frac{1}{2}v + 4N^2 x^2 cos^7 \frac{1}{2}v\, sin\frac{1}{2}v$$

and

$$\frac{sin\, g'' - sin\, g}{sin\, g\, cot(r - v)} = \psi = 2N\lambda\, cos^4 \frac{1}{2}v + 4N^2 \lambda^2 cos^7 \frac{1}{2}v\, sin\frac{1}{2}v.$$

Therefore, from these equations, the value of the number $N$ will be founded, from which the distance of the perihelion to the Sun $a$ becomes known, and the true anomaly[21] $v$ for the first observation, from which the time $k$, which the comet approaches the perihelion becomes known. Thereafter, from the knowledge of $v$, the angle $r$ becomes known, and then, finally, the longitude of the perihelion $p$. **Q.E.I.**

---

[21] See figure in Problem 3.

28. **Corollary 1.** Since we find out that

$$\varphi = \frac{\sin g' - \sin g}{\sin g \cot(r - v)} \quad \text{and} \quad \psi = \frac{\sin g'' - \sin g}{\sin g \cot(r - v)}$$

and that the angle $r - v$ is given by the equation $\sin(r - v) = \frac{\sin g}{\sin s}$, the decrement of the true anomaly $v$ in the second and the third observations can be found, that is, $\varphi$ and $\psi$.

29. **Corollary 2.** Then, since $\varphi$ and $\psi$ are now known, them from (23)

$$Nx = \frac{\varphi}{2\cos^4 \frac{1}{2}v} + \frac{\varphi^2 \sin \frac{1}{2}v}{2\cos^5 \frac{1}{2}v} + \text{etc.}$$

$$N\lambda = \frac{\psi}{2\cos^4 \frac{1}{2}v} + \frac{\psi^2 \sin \frac{1}{2}v}{2\cos^5 \frac{1}{2}v} + \text{etc.}$$

and then, by eliminating the number $N$ we will have that

$$\frac{\lambda}{x} = \frac{\psi \cos \frac{1}{2}v + \psi^2 \sin \frac{1}{2}v}{\varphi \cos \frac{1}{2}v + \varphi^2 \sin \frac{1}{2}v}, \quad \text{or} \quad \lambda\varphi + \lambda\varphi^2 \tan \frac{1}{2}v = x\psi + x\psi^2 \tan \frac{1}{2}v$$

from which it is quickly found the true anomaly $v$, because

$$\tan \frac{1}{2}v = \frac{x\psi - \lambda\varphi}{\lambda\varphi^2 - x\psi^2}.$$

30. **Corollary 3.** Then, once the angle $v$ has been found in this way, from it, the angle $r$ is immediately apparent, and hence, the longitude of the perihelion $p$ becomes known from the equations $\sin(r - v) = \frac{\sin g}{\sin s}$ and $\tan(p - q) = \tan r \cos s$. Then, truly, also the number $N$ is obtained from the equation

$$N = \frac{\varphi}{2x \cos^4 \frac{1}{2}v} + \frac{\varphi^2 \sin \frac{1}{2}v}{2x \cos^5 \frac{1}{2}v}$$

from which, the distance from the perihelion to the Sun $a$ can be determined as well.

31. **Scholium.** If the cosine of the angle $\frac{1}{2}v$ is rather small, the series by which the number N is defined converges very little, the sine calculation by approximation, after $\varphi$ and $\psi$ are found, can be established in the following way: since $Nk = \tan \frac{1}{2}v + \frac{1}{3}\tan^3 \frac{1}{2}v$, then

$$N(k - x) = \tan \frac{v - \varphi}{2} + \frac{1}{3}\tan^3 \frac{v - \varphi}{2}$$

and then,
$$Nx = \tan\frac{1}{2}v - \tan\frac{v-\varphi}{2} + \tan^3\frac{1}{2}v - \frac{1}{3}\tan^3\frac{v-\varphi}{2}.$$

Set $\tan\frac{1}{2}v = t$ and $\tan\frac{1}{2}\varphi = \mu$ as well as $\tan\frac{1}{2}\varphi = v$, and then, $\tan\frac{v-\varphi}{2} = \frac{t-\mu}{1+\mu t}$ and $\tan\frac{v-\psi}{2} = \frac{t-v}{1+vt}$, and once these are substituted results in

$$Nx = \frac{\mu(1+t^2)}{1+\mu t} + \frac{\mu t^2(1+t^2) - \mu^2 t(1-t^4) + \frac{1}{3}\mu^3(1+t^6)}{(1+\mu t)^3}, \text{ and}$$

$$N\lambda = \frac{v(1+t^2)}{1+vt} + \frac{vt^2(1+t^2) - v^2 t(1-t^4) + \frac{1}{3}v^3(1+t^6)}{(1+vt)^3}.$$

And by elimination N between these equations will result in this equation

$$\frac{\lambda(1+vt)^3}{x(1+\mu t)^3} = \frac{v(1+vt)^2 + vt^2 - v^2t(1-t^2) + \frac{1}{3}v^3(1-t^2+t^4)}{\mu(1+\mu t)^2 + \mu t^2 - \mu^2 t(1-t^2) + \frac{1}{3}\mu^3(1-t^2+t^4)}, \text{ or}$$

$$\frac{\lambda(1+vt)^3}{x(1+\mu t)^3} = \frac{v + v^2 t + \frac{1}{3}v^3(1+t^2)}{\mu + \mu^2 t + \frac{1}{3}\mu^3(1+t^2)}.$$

32. **Scholium 2.** Although we may find only approximate values for $\varphi$ and $\psi$ only by approximation, assuming three observations near to one another, although they can also be found exactly, even if the observations are utmost far away from each other. Having been found $q$ and $s$ by the prescribed way, which did not lean upon in any approximation, the angle $r - v$ becomes readily known, since, as before

$$\sin(r - v) = \frac{\sin g}{\sin s}, \quad \text{or} \quad \tan(r - v) = \frac{\tan(f - q)}{\cos s}.$$

Then, the angle $\varphi$ becomes also known from the equation $\sin(r - v + \varphi) = \frac{\sin g'}{\sin s}$, and the angle $\psi$ from the equation $\sin(r - v + \psi) = \frac{\sin g''}{\sin s}$. Hence, having been found $\varphi$ and $\psi$, then, by the method exhibited in the preceding Scholium, $N$ and $v$ will be discovered. Therefore, from any three locations of heliocentric comets, that is, from the longitudes and latitudes, the orbit of the comet can be exactly determined. However, before this method is adopted to determine with great precision the orbit of a comet, from three observations near to one another, and at the same time far away from the perihelion, provided that equation of order five above can be easily solved. For cases where $\mu$ and $v$ are made quite small, then as approximations

$$\frac{\lambda(1+vt)^3}{x(1+\mu t)^3} = \frac{v}{\mu} \quad \text{and} \quad \frac{1+vt}{1+\mu t} = \frac{\sqrt[3]{xv}}{\sqrt[3]{\lambda\mu}}, \quad \text{whence} \quad t = \frac{\sqrt[3]{\lambda\mu} - \sqrt[3]{xv}}{\mu\sqrt[3]{xv} - v\sqrt[3]{\lambda\mu}} ;$$

from this approximate value, a more accurate value can be easily drawn.